\documentclass[10pt,letterpaper,twocolumn]{article}
\usepackage[T1]{fontenc}
\usepackage{lmodern}
\usepackage[letterpaper,top=.72in,bottom=.74in,left=.76in,right=.76in,columnsep=.26in]{geometry}
\usepackage{microtype,graphicx,amsmath,amssymb,booktabs,array}
\newlength{\interactionpanelwidth}
\usepackage[round,authoryear]{natbib}
\setcitestyle{aysep={,},yysep={,},notesep={; }}
\usepackage[font=small,labelfont=bf]{caption}
\usepackage{placeins,flushend}
\usepackage[hidelinks]{hyperref}
\hypersetup{pdftitle={From Sketch to Design: A Cross-scale Workflow for Procedural Generative Urban Design},pdfauthor={Jin Gao and Sayjel Vijay Patel},pdfkeywords={generative urban design, procedural modeling, urban semiotics, gesture recognition, tensor fields, creativity support tools}}
\newcommand{\figref}[1]{\mbox{Figure~\ref{fig:#1}}}
\begin{document}
\twocolumn[{\begin{minipage}{\textwidth}
\centering
{\LARGE\bfseries From Sketch to Design\par}
\vspace{3pt}
{\large A Cross-scale Workflow for Procedural Generative Urban Design\par}
\vspace{9pt}
{\large Jin Gao\textsuperscript{1} and Sayjel Vijay Patel\textsuperscript{2}\par}
\vspace{4pt}
{\small\textsuperscript{1}Massachusetts Institute of Technology\quad\textsuperscript{2}Digital Blue Foam\par
\href{mailto:gaojin@mit.edu}{gaojin@mit.edu}\quad\href{mailto:sayjel@digitalbluefoam.com}{sayjel@digitalbluefoam.com}\par
Jin Gao: \href{https://orcid.org/0000-0002-1595-6895}{ORCID 0000-0002-1595-6895}\par}
\vspace{7pt}
{\small Author version of the paper published in \emph{Accelerated Design: CAADRIA 2024},\par
Volume 1, pp.~343--352. \href{https://doi.org/10.52842/conf.caadria.2024.1.343}{doi:10.52842/conf.caadria.2024.1.343}\par}
\vspace{10pt}
\begin{minipage}{.96\textwidth}\small
\textbf{Abstract.} Urban designers use sketches and symbols to express spatial intentions, while many generative urban design tools rely on numeric parameters that are difficult to relate to this visual process. We propose a cross-scale workflow and a web-based prototype that translates symbolic sketches into controllable procedural urban generation. A vocabulary of shapes, informed by examples of urban design diagrams and sketching exercises with graduate urban design students, encodes attributes such as network structure, density, and program. The prototype uses the \$1 Unistroke Recognizer to map gestures to symbols and translates their locations and sizes into tensor-field attractors and weighted parameter grids. Grid and radial tensor fields guide street-network generation, while the parameter grids control building heights and programs. Implemented in JavaScript and TypeScript with React and Three.js, the prototype supports interactive sketching, visualization, and urban massing generation. We describe the methodology and illustrate prototype interactions, then discuss downstream integration with architectural-scale generative tools for further refinement and analysis. The work connects designers' symbolic input with an explicit procedural generation process.

\textbf{Keywords:} generative urban design, procedural modeling, urban semiotics, gesture recognition, tensor fields, creativity support tools
\end{minipage}
\vspace{10pt}
\end{minipage}\vspace{10pt}}]
\section{Introduction}\label{introduction}

Urban designers use diagrams and symbols to reason about spatial relationships and communicate intentions about density, form, and land use. Theories of urban perception and meaning provide a context for these representations. In \emph{The Image of the City}, \citet{lynch1964} examines how recurring urban elements contribute to legible mental images of cities. \citet{zhu1993} approaches urban design through spatial and symbolic relationships. These perspectives motivate our use of sketches as inputs through which designers can express spatial intentions.

Procedural urban modeling offers several ways to encode spatial structure, including grammar-based design patterns and interactively controlled street networks \citep{beirao2010,chen2008}. However, translating a design intention into generation parameters remains an interface problem. Parameters may be entered through numeric controls, derived from GIS data, or specified through spatial constraints. These inputs do not necessarily correspond directly to the symbols that designers use in sketches. Our work addresses this translation by connecting graphical inputs to explicit procedural controls and visual feedback.

Existing tools already support interactive urban design, including tangible interfaces and parameterized generation workflows \citep{ishii2002,koenig2017,alonso2018}. Their input methods and operating environments differ, however. In conventional modeling workflows, designers may need to manipulate many urban masses individually, while generative workflows require them to express their intentions through the controls provided by the tool. Our goal is to support rapid sketch-based modification within a browser interface that can connect to downstream architectural design workflows.

The objective of this study is to propose a cross-platform generative urban design tool that simplifies urban design workflow. This tool aims to bridge the gap between core urban design methodologies and the intuitive, creative process of designers. The significance of this study lies in its potential to make urban design more accessible, efficient, and integrated with existing design workflows.

\section{Related Work}\label{related-work}

\subsection{Vocabulary of Urban Design}\label{vocabulary-of-urban-design}

Architectural and urban design can be organized through reusable vocabularies of spatial relationships. \citet{alexander1977} describe interconnected patterns spanning settlements, buildings, and construction, while \citet{white1975} uses diagrams to explore architectural form and design concepts. We draw on this use of visual vocabularies to define a compact set of sketch symbols for the prototype.

\subsection{Generative Urban Design}\label{generative-urban-design}

Urban generation methods differ in the processes they represent and the controls they expose. \citet{batty1997} describes how local cellular-automaton rules can produce complex urban patterns. In their review of urban growth prediction, \citet{triant2012} discuss cellular-automaton approaches, including SLEUTH, whose inputs include slope, land use, exclusion, urban extent, transportation, and hillshade. Such models describe growth dynamics through local transition rules.

Other approaches encode design decisions more directly. \citet{beirao2010} implement grammar-based urban design patterns in an AutoCAD Civil 3D prototype. \citet{mei2021} translate planning regulations and density controls into urban form across block and district scales. \citet{wilson2019} combine procedural geometry generation with performance evaluation and communication to support the comparison of large sets of masterplan alternatives.

A related line of work examines the representations through which designers engage with urban data. \citet{sandoval2016} proposes \emph{Painting with Data}, combining voxel-based representation with a graph database. \citet{sandoval2021} subsequently examines the histories of geospatial computing and the role of drawing and computation in collaborative planning and design. These works motivate attention to interfaces and representations alongside generation algorithms.

Machine learning provides another approach to urban form generation. \citet{boim2022} train generative adversarial networks on patterns of non-planned settlements to generate development alternatives with morphological similarities to the training fabric. Our prototype instead uses explicit procedural fields so that sketch symbols can be mapped to interpretable generation controls.

For street generation, we build on the interactive procedural framework of \citet{chen2008} and its open-source TypeScript implementation, MapGenerator \citep{keir2023}. Grid and radial tensor fields describe preferred local street orientations; constraints and field editing can adapt those orientations to site boundaries. We use this framework as the generation component of our interface and add a control layer that translates recognized sketches into field attractors and urban parameters.

This paper examines three questions:

\begin{itemize}
\item
  How do symbols function as a fundamental language in complex urban design proposals, and how can they be defined and utilized in the design process?
\item
  What are the challenges and variations in parameters controlling procedural urban modeling, and how can a generative urban design tool address them to align with the visual thinking process while providing efficient and adaptable solutions?
\item
  What are the current limitations and challenges in existing interactive urban design tools, particularly in relation to their understanding of the symbolic language used by designers, and how can a cross-platform generative urban design tool bridge the gap between traditional urban design workflow and the creative design process?
\end{itemize}

\subsection{Design Tools and Human-Computer Interaction}\label{design-tools-and-human-computer-interaction}

Urban design software includes both plugins within modeling environments and standalone applications. \citet{koenig2017} present Grasshopper/Rhino components for synthesizing street networks, parcels, and building volumes. Urban Network Analysis is available in both ArcGIS and Rhino: \citet{sevtsuk2012} describe the ArcGIS toolbox, while \citet{sevtsuk2016} document its Rhinoceros~3D implementation for network-based accessibility analysis. \citet{kelly2021} describes CityEngine's rule-based pipeline for transforming two-dimensional data into three-dimensional urban models. These examples illustrate different combinations of generation, analysis, and host-platform integration.

Tangible systems offer a different approach to interaction. The Augmented Urban Planning Workbench of \citet{ishii2002} overlays drawings, physical models, and projected simulation in a shared workspace. CityScope \citep{alonso2018} combines a tangible interface with analysis and visualization of proposed urban interventions to support stakeholder discussion. These projects demonstrate ways to connect spatial manipulation with computational feedback. Our prototype explores browser-based sketch interaction as another way to make generation controls directly accessible to designers.

Web-based tools provide further examples of interfaces for urban design and scenario exploration. Digital Blue Foam provides a generative design platform \citep{dbf2023}. \citet{delve2023} describe the neighborhood-generation workflow that subsequently became Delve. KPF Urban Interface's Scout provides a web interface for visually exploring design alternatives and associated data \citep{scout2023}. These examples provide context for our investigation of symbolic sketch input as a way to control procedural urban generation.

\section{Methodology}\label{methodology}

\subsection{Defining a Gesture Language System}\label{defining-a-gesture-language-system}

\begin{figure}[!ht]
\centering
\includegraphics[width=\linewidth]{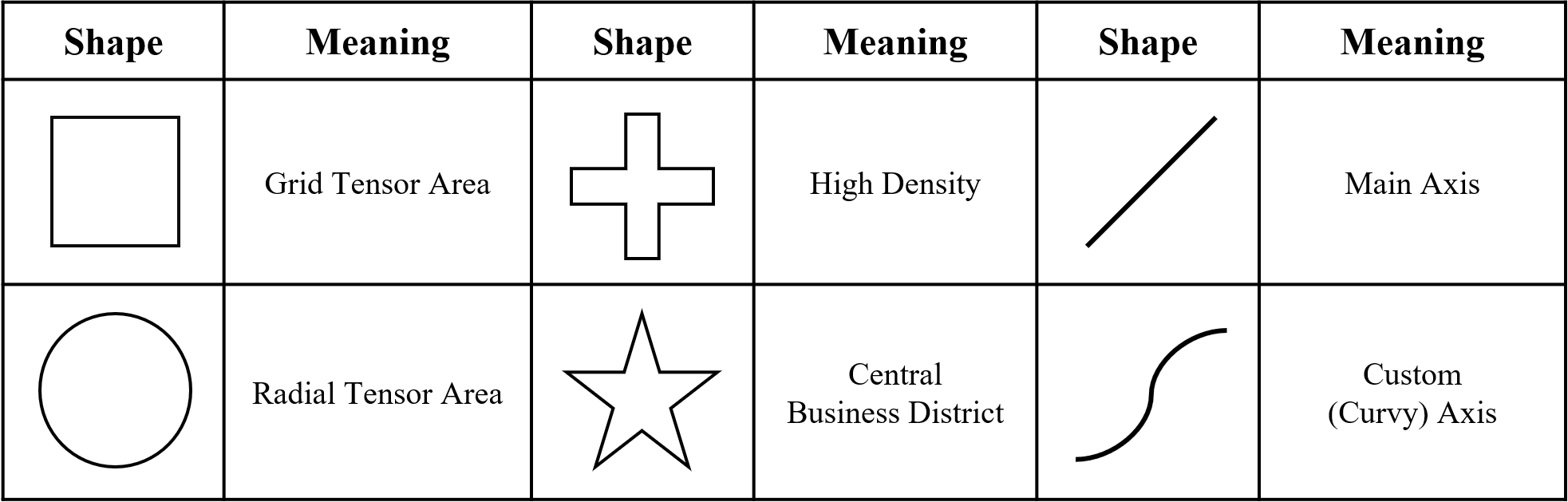}
\caption{Examples of the urban design symbolic language.}
\label{fig:1}
\end{figure}

To understand how designers utilize symbolic language in their sketches, we collected and analysed samples of urban design diagrams, and conducted experiments under a controlled setting by asking graduate-level urban design major students to draw interpretations of specific sites using their own symbolic language. We then categorized commonly used symbolic languages into basic shapes like stars, squares, and circles, to define a system of symbols. As samples listed in \figref{1}, each of these symbols is meant to represent different types of influential parameters in urban design, including height, density, and program, which eventually shape the physical and functional aspects of the generated urban spaces.

\subsection{Cross-scale Workflow}\label{cross-scale-workflow}

\begin{figure*}[!t]
\centering
\includegraphics[width=\linewidth]{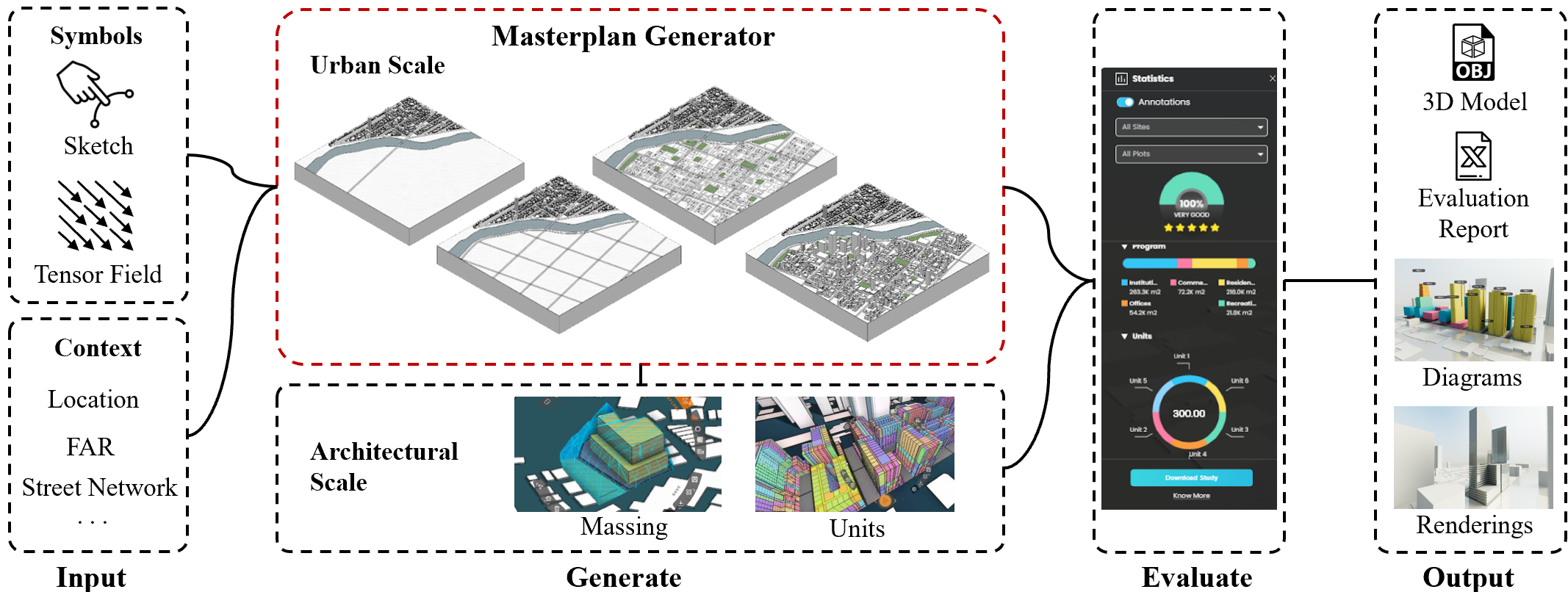}
\caption{Proposed cross-scale generative urban design workflow. The prototype focuses on urban-scale generation; architectural refinement and analysis indicate downstream integration.}
\label{fig:2}
\end{figure*}

The research proposes a cross-scale workflow for generative urban design (\figref{2}). The workflow takes designers' sketches and context information as inputs and generates urban-scale massing. It envisages subsequent refinement at the plot and building scales, including architectural massing and room division, through downstream generative design tools. The proposed analysis stage includes urban-scale measures such as proximity, accessibility, porosity, and floor area ratio (FAR), together with architectural measures such as floor area, structure, material usage, and energy efficiency. Standard 3D model outputs would support detailed rendering and analysis in dedicated professional software. The prototype reported here focuses on urban masterplan and massing generation; architectural refinement and comprehensive performance analysis are directions for further integration.

\subsection{Prototype Implementation}\label{prototype-implementation}

\subsubsection{Software Architecture}\label{software-architecture}

\begin{figure*}[!t]
\centering
\includegraphics[width=\linewidth]{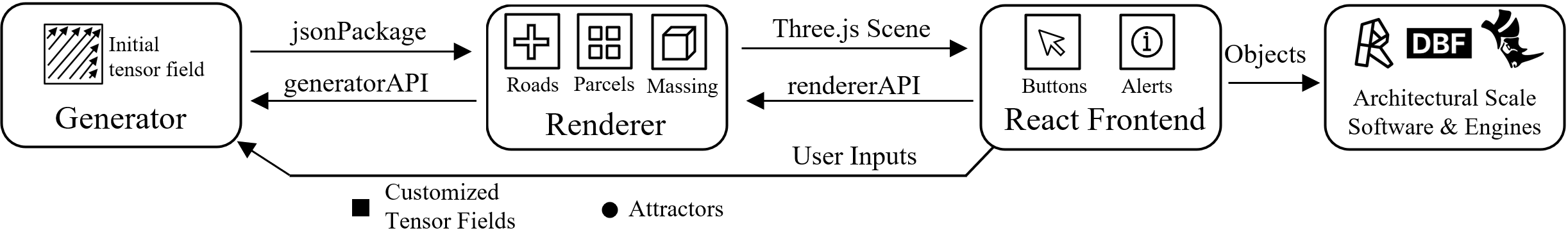}
\caption{Software architecture: generation, rendering, and frontend components exchange JSON data through internal APIs.}
\label{fig:3}
\end{figure*}

The prototype is a web application that mainly features urban masterplan generation. Shown in \figref{3}, it consists of three components, a generation engine running at the backend, a render engine to visualize the generated models, and a frontend framework taking users' inputs. Between each component, the information is interchanged through internal APIs in the format of JSON objects.

\subsubsection{Symbolic Inputs and Gesture Recognition}\label{symbolic-inputs-and-gesture-recognition}

The prototype uses the \$1 Unistroke Recognizer introduced by \citet{wobbrock2007}. This instance-based nearest-neighbor classifier matches normalized gestures to geometric templates. In our interface, recognized symbols are translated into attractors with a position and size. Users can switch between graphical tools such as pencils, erasers, and lines. The resulting symbolic inputs update a multi-dimensional parameter grid that influences the generative process.

\subsubsection{Urban Environment Generation}\label{urban-environment-generation}

We selected a tensor-field approach because it exposes spatial controls that can be connected to sketches. The prototype incorporates the urban generation module of MapGenerator \citep{keir2023}, a TypeScript implementation used here with the street-modeling approach of \citet{chen2008}. We add a control layer that converts user sketches into attractors and grid parameters. The prototype updates its urban geometry from the resulting field and parameter grid.

\textbf{Step 1: Environment Definition.} Within a set rectangle boundary with a range of \(1000\,\mathrm{m} \times 1000\,\mathrm{m}\) (can be modified), the algorithm first randomly generates a coastline that defines the boundary between land and water. The boundary serves as the base input for the context of the urban layout and influences the overall structure of the tensor field, therefore influencing the generated urban network. For $R\geq0$ and $\theta\in[0,2\pi)$, a symmetric traceless tensor has the representation given by \citet{chen2008}:

\begin{equation} T=R\begin{bmatrix}\cos 2\theta & \sin 2\theta \\ \sin 2\theta & -\cos 2\theta\end{bmatrix}. \label{eq:tensor}\end{equation}

\textbf{Step 2: Urban Network Generation.} The street network is modeled as a graph \(G=(V,E)\), where \(V\) is the set of nodes (intersections and endpoints) and \(E\) is the set of edges (streets). Initially, the site has a default grid tensor field. Users can introduce attractors by adding symbols. The attractors represent different characteristics of the basis tensor field. Currently, the basis tensor field has grid and radial types. The algorithm computes gradients between each tensor field defined by the attractor and updates dynamically. Following the basis-field construction and spatial weighting of \citet{chen2008}, we express the local grid contribution below. For a direction vector $v=(v_x,v_y)$, the magnitude and angle are:

\begin{equation}
\ell=\sqrt{v_x^2+v_y^2},\qquad \theta=\arctan\!\left(\frac{v_y}{v_x}\right).
\end{equation} \begin{equation}
T_{\mathrm{grid}}(p)=e^{-d\lVert p-p_0\rVert^2}\begin{bmatrix}\cos 2\theta & \sin 2\theta \\ \sin 2\theta & -\cos 2\theta\end{bmatrix}.\label{eq:grid}
\end{equation}

\begin{minipage}{\columnwidth}
The corresponding spatially weighted radial contribution is:

\begin{equation}
T_{\mathrm{radial}}(p)=e^{-d\lVert p-p_0\rVert^2}
\begin{bmatrix}y^2-x^2&-2xy\\-2xy&-(y^2-x^2)\end{bmatrix},\label{eq:radial}
\end{equation}

where \(x=x_p-x_0\) and \(y=y_p-y_0\). Here \(p_0\) denotes the attractor location, \(d\) is a decay parameter, and \(v=(v_x,v_y)\) specifies the grid direction.
\end{minipage}

\begin{figure}[!t]
\centering
\includegraphics[width=\linewidth,trim=0 56.25bp 0 0,clip]{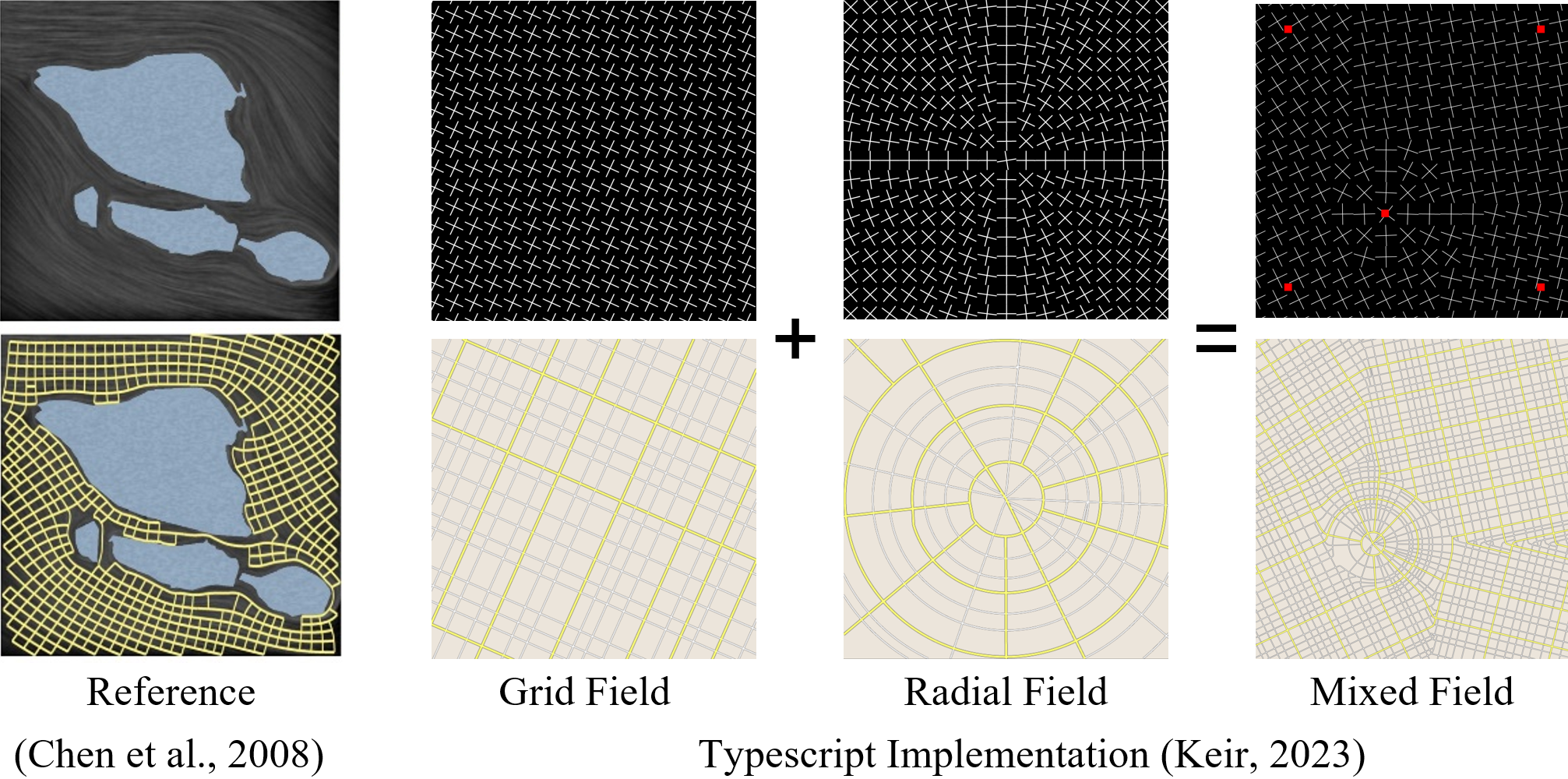}
\par\vspace{2pt}
{\scriptsize\begin{tabular*}{\linewidth}{@{\extracolsep{\fill}}cccc@{}}
Reference & Grid field & Radial field & Mixed field
\end{tabular*}}
\caption{Tensor fields and generated street graphs. Left reference panel: \citet{chen2008}. Grid, radial, and mixed-field examples: MapGenerator \citep{keir2023}.}
\label{fig:4}
\end{figure}

\figref{4} illustrates how the algorithm finds street networks as streamlines within tensor fields. The street network is then divided into two hierarchies: major and minor roads. The hierarchy is defined by the distance and numbers between streamlines.

\begin{figure*}[!t]
\centering
\includegraphics[width=0.96\linewidth]{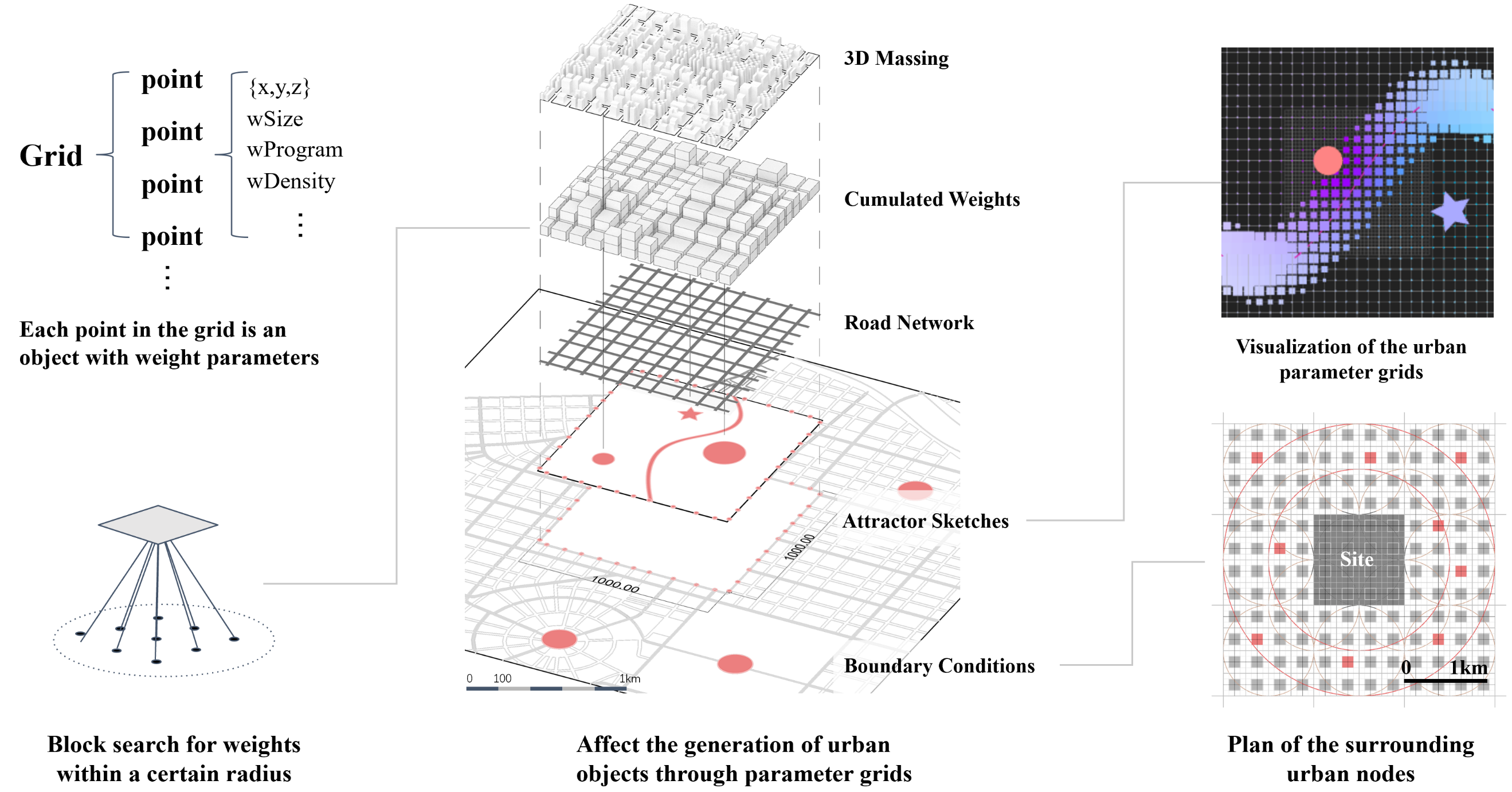}
\caption{Symbolic attractors and weighted parameter grids guide 3D massing generation.}
\label{fig:5}
\end{figure*}

\textbf{Step 3: 3D Massing Generation.} Shown in \figref{5}, the 3D massing is further defined by a grid system that incorporates various weighted parameters based on user inputs and surrounding contexts. For instance, a ``star'' shape is interpreted as representing a ``central area''. It will have a high influence on the density and program score of the grid parameters around its central point. Then, based on the combined influences of scores, the building base plane will be assigned a weighted height and program parameter that will be represented by the final massing outcomes. Currently, the massing is extruded from the base planes. For further development, the massing and its corresponding parameters can be exported to architectural scale generative design tools, allowing further massing division and detailed architectural design.

\begin{figure*}[!t]
\centering
\setlength{\interactionpanelwidth}{0.324\linewidth}
\includegraphics[width=\interactionpanelwidth,trim=0 44.25bp 941bp 154.875bp,clip]{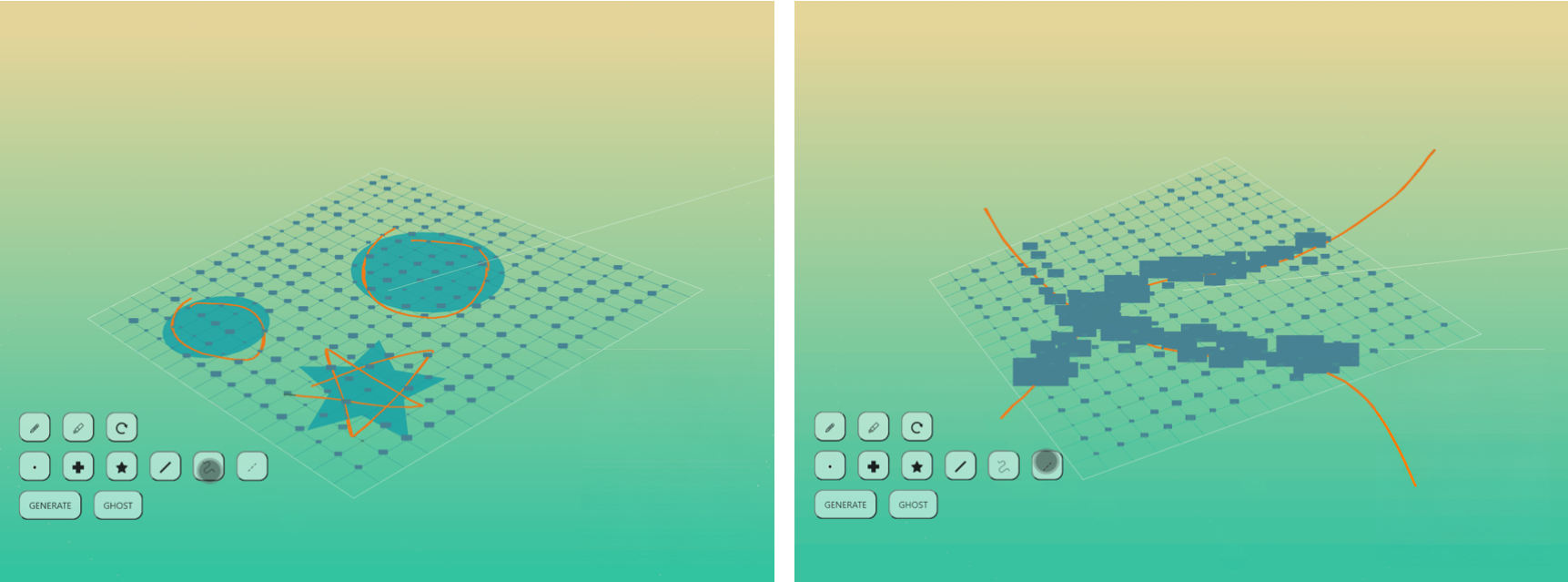}\hfill
\includegraphics[width=\interactionpanelwidth,trim=873bp 44.25bp 68bp 154.875bp,clip]{figures/image12-no-stats.png}\hfill
\includegraphics[width=\interactionpanelwidth,trim=0 102.31875bp 181.9bp 95.8125bp,clip]{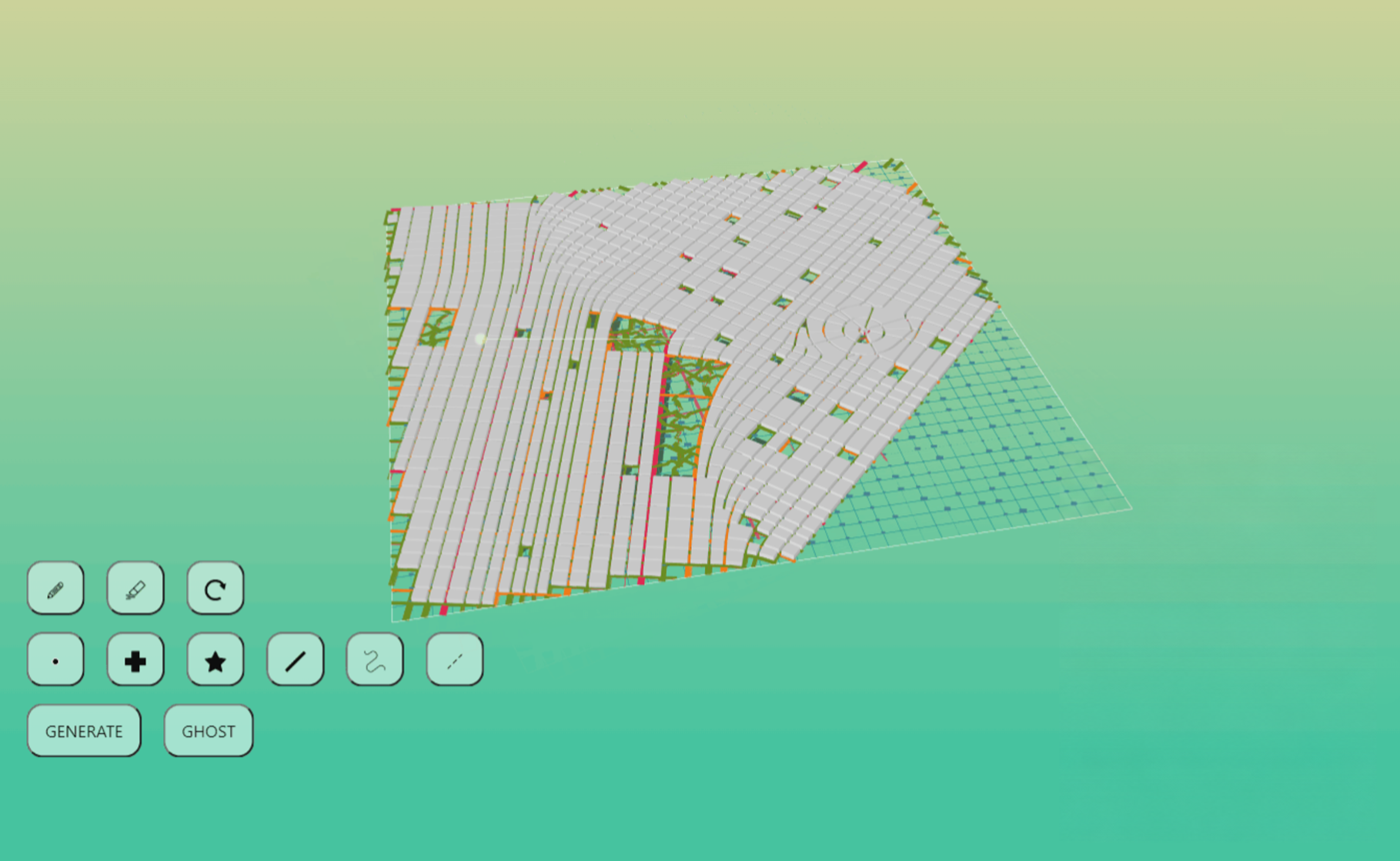}\\[4pt]
\includegraphics[width=0.494\linewidth]{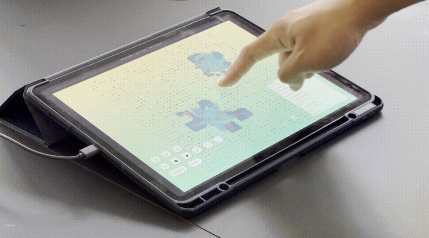}\hfill
\includegraphics[width=0.494\linewidth]{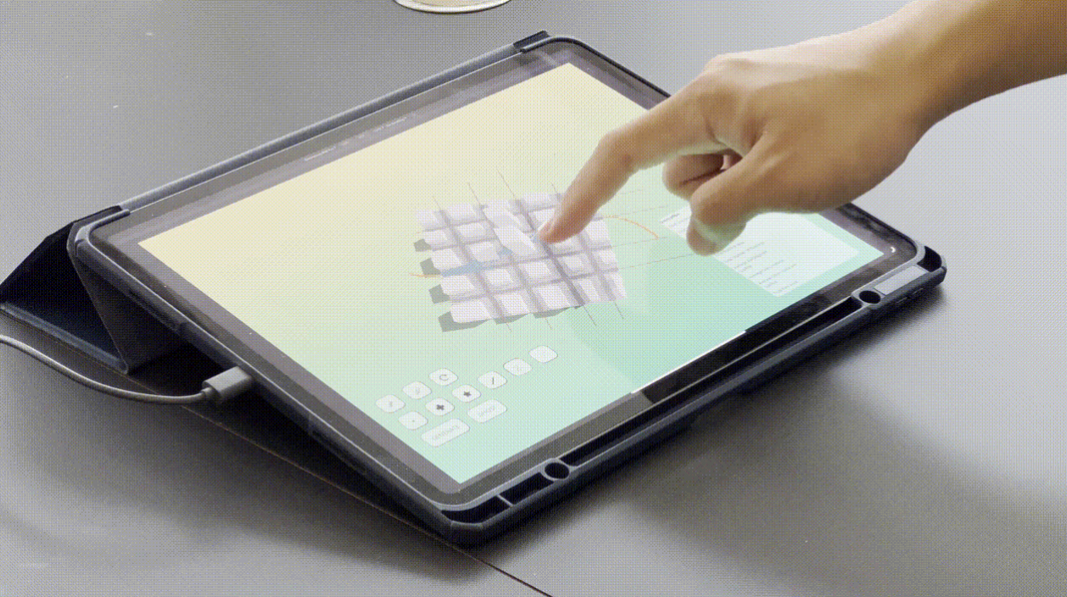}
\caption{Prototype interaction examples: symbolic sketch inputs and generated massing in the web interface (top), and touch-based interaction on a tablet (bottom). Statistics panels have been removed for visual clarity.}
\label{fig:6}
\end{figure*}

\textbf{Step 4: Visualization and Representation.} The prototype's visualization component is powered by Three.js \citep{threejs2023}, a widely used WebGL library, to achieve a high-performance real-time rendering of the generated urban environment. The React framework is used to create a control layer that receives user sketches, with the dat.GUI as general parameter input panel. The program provides augmented representation by adjusting material properties based on inherent urban parameters. Furthermore, the user can switch between display modes (perspective and isometric) and perform real-time rendering using additional libraries like three-gpu-pathtracer \citep{gkjohnson2023}. \figref{6} shows the live interactions a user creates with the prototype.

\textbf{Step 5: Post-process of Generation Results.} The generation results are designed to be exported as compatible formats for architectural scale generative software, ensuring further manipulation and refined design by the next step of workflows. Additionally, the model can be exported as a standalone entity and imported into common modeling software utilized as the foundational model for proposal graphics and further rendering.

\section{Discussion and Future Work}\label{discussion-and-future-work}

\subsection{Discussion}\label{discussion}

Unlike traditional urban design tools or modeling software that often rely on intricate inputs and non-dedicated workflow for urban designers, this tool proposes an intuitive approach by interpreting designers' symbolic languages and sketches into the computer-aided urban design workflow. The tool stresses interactivity, which is fundamental in fitting the actual needs of designers. Currently, with more advanced architecture-level generative design tools coming out, the research aims at filling the gap in urban-scale design tools and seeks to support a transition from urban-scale design to detailed architectural design. Potential use cases include helping urban planners and designers to quickly conceptualize and manipulate urban layouts, and preview the outcome in a cross-platform, interactive user interface. In academic settings, it can be used as a teaching tool to demonstrate key principles of urban design. Students could use the tool to experiment with multiple design approaches. The tool could also contribute to participatory design by allowing non-professionals to contribute to design discussions in a more interactive and intuitive approach.

\subsection{Future Work}\label{future-work}

As the prototype is still in its development phase, there are several key areas of enhancement planned for the next phase to make the proposed workflow more robust and adaptable. First, import real-world map data as context, to provide an adaptive urban network based on surrounding environments. Defining more types of tensor fields will make the tool more adaptable in representing multiple typologies of urban networks. Furthermore, by expanding and customizing the symbol language system, we can make the tool adapt to a broad range of users. We foresee opportunities below:

\begin{itemize}
\item
  Seamless cross-scale design tool: Currently, the prototype is a middleware between the user's input and architectural scale design software. This barrier can be broken by establishing dynamic links and automatic sync between tools.
\item
  Collaborative symbol language-aided design: Like real-world design practice, there are usually several people sketching on the same piece of masterplan together. With the collaborative editing feature, multiple designers or community members can discuss and contribute to the design outcome under a shared platform.
\item
  Multimodal symbol-driven modeling: The emergence of large language models provides powerful tools for understanding human language. Following the principle of this paper, we can categorize the designer's linguistic symbol vocabularies and integrate them with the workflow, to achieve the connection between language and interactive modeling. Moreover, emerging visualization technologies like augmented and virtual reality enable spatial symbol input and an immersive preview of the results. We believe its corresponding workflow will open opportunities for future urban planning and design with further possibilities for creative exploration.
\end{itemize}

\section*{Acknowledgments}

The authors would like to thank the Digital Blue Foam team for their generous help and guidance during the process of development of the prototype.

{\small\raggedright
}
\end{document}